\documentclass[11pt]{article}

\usepackage{geometry}
\usepackage{currfile}
\usepackage{textcomp}

\usepackage{url}
\usepackage{doi}

\usepackage{xcolor,soul}
\sethlcolor{yellow}

\usepackage[pdftex]{graphicx}
\DeclareGraphicsExtensions{.pdf,.jpeg,.png}
\graphicspath{{./.}}

\date{\today}

\begin{document}


\title{\Large{Trust Infrastructures for Virtual Asset Service Providers\\
~~\\
{\large (Extended Abstract)} }   \\
~~}
\author{
\large{Thomas~Hardjono}\\
\large{~~}\\
\large{MIT Connection Science \& Engineering}\\
\large{Massachusetts Institute of Technology}\\
\large{Cambridge, MA 02139, USA}\\
\large{~~}\\
\small{{\tt hardjono@mit.edu}}\\
\large{~~}\\
}

\maketitle

\begin{abstract}
Virtual asset service providers (VASPs) currently face a number of challenges,
both from the technological and the regulatory perspectives.
In the context of virtual asset transfers
one key issue is the need for VASPs to
securely exchange customer information to comply to the Travel Rule.
We discuss a VASP {\em information sharing network}
as one form of a trust infrastructure for VASP-to-VASP interactions.
Related to this is the need for a trusted identity infrastructure for VASPs
that would permit other entities to quickly ascertain the legal business status of a VASP.
For regulated wallets, an attestation infrastructure may provide VASPs
and insurance providers with better visibility into the state of wallets based on
trusted hardware.
Finally, for customers of VASPs
there is a need for seamless integration between the VASP services 
with the existing consumer identity management infrastructure,
providing a user-friendly experience for transferring virtual assets
to other users.\\
~~\\
Keywords: Virtual Assets, Blockchains, Trust, Infrastructure
\end{abstract}

\newpage
\clearpage


{\small 
\tableofcontents
}

\newpage
\clearpage



\section{Introduction}

It has been over a decade since the advent of the Bitcoin cryptocurrency system~\cite{Bitcoin}
based on the hash-chain model of Haber and Stornetta~\cite{HaberStornetta1991}.
Considerable interest, hype and speculative investments have gone
into various projects on blockchains
aimed at developing decentralized virtual asset ecosystems.
At the same time,
there has been a growing trend of theft and 
misappropriation of virtual assets~\cite{Su2019-Forbes,Reddy2019}.
Additionally,
crypto-currencies are still being exploited for 
money-laundering~\cite{SchoenbergRobinson2018,Canellis2019},
raising significant concern on the part of regulators.

The FATF Recommendation {No.~15}~\cite{FATF-Recommendation15-2018}
that was finalized in mid-2019
provided an unambiguous definition of virtual assets  
and virtual asset service providers (VASP).
This places VASPs under much of the same AML-related
regulations as that of traditional financial institutions,
notably the Travel Rule.
In short,
the Travel Rule requires a VASP to obtain, validate and retain 
their customer's personal information need for compliance.
In the case of virtual asset transfers between an originator and a beneficiary,
their corresponding VASPs must share peer-wise the customer information
pertaining to the originator and beneficiary.

Today many users possessing virtual assets (e.g. cryptocurrencies)
expect asset transfers through VASPs
to be confirmed or settled in a matter of seconds.
However, the need for VASPs to exchange and validate customer information prior to
asset transfers may impose delays on the settlement of transfers.
Furthermore, 
VASPs do not as yet have an agreed mechanism to exchange
their respective customer information in a secure and reliable manner.

We believe that this lack of an information exchange mechanism
points to a more fundamental challenge facing the VASP community worldwide:
namely the lack of {\em trust infrastructures}
that are highly scalable and interoperable,
which permit business-trust and legal-trust to be established 
for peer-to-peer transactions of virtual assets
across different jurisdictions.

There are several forms of trust infrastructures needed for VASPs,
and in the current work we discuss three forms or types of such infrastructures.
The first is an {\em information sharing infrastructure} 
specifically for VASPs.
The main purpose of a VASP information sharing infrastructure 
is to securely and confidentially share customer
information related to transfers of virtual assets.
Related to this network is the VASP {\em identity infrastructure}
that permits VASPs and other entities to quickly ascertain
the business legal status of other VASPs.
Next is an {\em attestation infrastructure} that 
can support VASPs and asset insurers
in obtaining better
visibility into the state of customer wallets based on
trusted hardware.
Finally, there is a need for a {\em claims infrastructure} for customer data sources
that integrates seamlessly into
the existing user digital identity infrastructure.

\section{The Travel Rule and VASP Customer Information}
\label{sec:TravelRule}

With the emergence of blockchain technologies, virtual assets and cryptocurrencies,
the FATF recognized the need to adequately mitigate the money laundering
(ML) and terrorist financing (TF) risks associated with virtual asset activities.
The {\em Financial Action Task Force} (FATF) is an inter-governmental body 
established in 1989 by the ministers of its member countries or jurisdictions.  
The objectives of the FATF are to set standards and promote effective 
implementation of legal, regulatory and operational measures 
for combating money laundering, terrorist financing and other 
related threats to the integrity of the international financial system.


The FATF Recommendation~15~\cite{FATF-Recommendation15-2018,FATF-Guidance-2019}
defines a {\em virtual asset} as 
a digital representation of value that can be 
digitally traded, or transferred, and can be used for payment or investment purposes. 
Virtual assets do not include digital representations of fiat currencies, 
securities and other financial assets that are already covered elsewhere in the FATF Recommendations.
The FATF defines a {\em virtual asset service provider} (VASP) to be 
any natural or legal person who is not covered elsewhere under the Recommendations, 
and as a business conducts one or more of the following activities or 
operations for or on behalf of another natural or legal person:
(i) exchange between virtual assets and fiat currencies; 
(ii) exchange between one or more forms of virtual assets;
(iii) transfer of virtual assets;
(iv) safekeeping and/or administration of virtual assets or instruments enabling control over virtual assets; and
(v) participation in and provision of financial services related to an issuer's offer and/or sale of a virtual asset.

The implication of the Recommendation~15, among others,
is that VASPs must be able to share the
originator and beneficiary information for virtual asset transactions.
This process -- also known as the {\em Travel Rule} --
originates from under the US Bank Secrecy Act (BSA - 31 USC 5311 - 5330),
which mandates that financial institutions deliver certain types of information
to the next financial institution when a funds transmittal event 
involves more than one financial institution.
This customer information includes (i) originator's name;
(ii) originator's account number (e.g. at the Originating-VASP);
(iii) originator's geographical address, or national identity number, 
or customer identification number (or date and place of birth);
(iv) beneficiary's name;
(v) beneficiary account number (e.g. at the Beneficiary-VASP).


It is important to emphasize that a VASP as a business entity
must be able to respond comprehensively to legitimate
inquiries from law enforcement
regarding one or more of its customers owning virtual assets
(e.g. legal SAR inquiries/warrants).
More specifically,
both the Originator-VASP and Beneficiary-VASP must possess 
the complete and accurate {\em actual} personal information (i.e. data)
regarding their account holders (i.e. customer).
This need for actual data, therefore, precludes the use of advanced cryptographic techniques
that aim to prevent disclosure while yielding implied knowledge,
such as those based on Zero-Knowledge Proof (ZKP) schemes~\cite{GoldwasserMicali1986-ZKP}.


Although beyond the scope for the current work,
one of the key challenging issues related to the Travel Rule
is the privacy of customer information 
once it has been delivered between VASPs.
This problem can be acute when one VASP is located within
a jurisdiction with strong privacy regulations (e.g. EU with GDPR~\cite{GDPR}),
while the other is located in a jurisdiction with an incompatible 
privacy regulations~\cite{FATF-12MonthReview-2020}.
More specifically,
if a Beneficiary-VASP is located under a different legal jurisdiction
(e.g. foreign country) observing weaker privacy regulations than the originator's jurisdiction,
there are no means for the originator to ensure her customer information
is not leaked or stolen from that Beneficiary-VASP.

\section{Information Sharing Infrastructure for VASPs}
\label{sec:GlobalVASPMessagingNetwork}

A core part of an information sharing infrastructure
is a network shared among VASPs to exchange information 
about themselves and their customers.
The notion of an {\em out-of-band} (off-chain) network
for VASPs to share information about themselves and their customers
was first proposed in~\cite{Hardjono2019b} as part of the 
broader discussion within the FATF Private Sector Consultative Forum
leading up to the finalization of Recommendation~15 in mid-2019.
The idea of an information sharing network is not new,
and the banking community established a similar network
(i.e. the SWIFT network~\cite{SwiftNet2004}) over two decades ago.
Today this network is the backbone for global correspondent banking.

As such, similar to banking data networks,
a ``network'' is needed for VASPs to securely exchange information 
about themselves and about their customers
for Travel Rule compliance and other requirements.
This information sharing network should be layered atop the proven TCP/IP Internet,
in order to provide the best connection resilience and speed.

\begin{figure}[t]
\centering
\includegraphics[width=1.0\textwidth, trim={0.0cm 0.0cm 0.0cm 0.0cm}, clip]{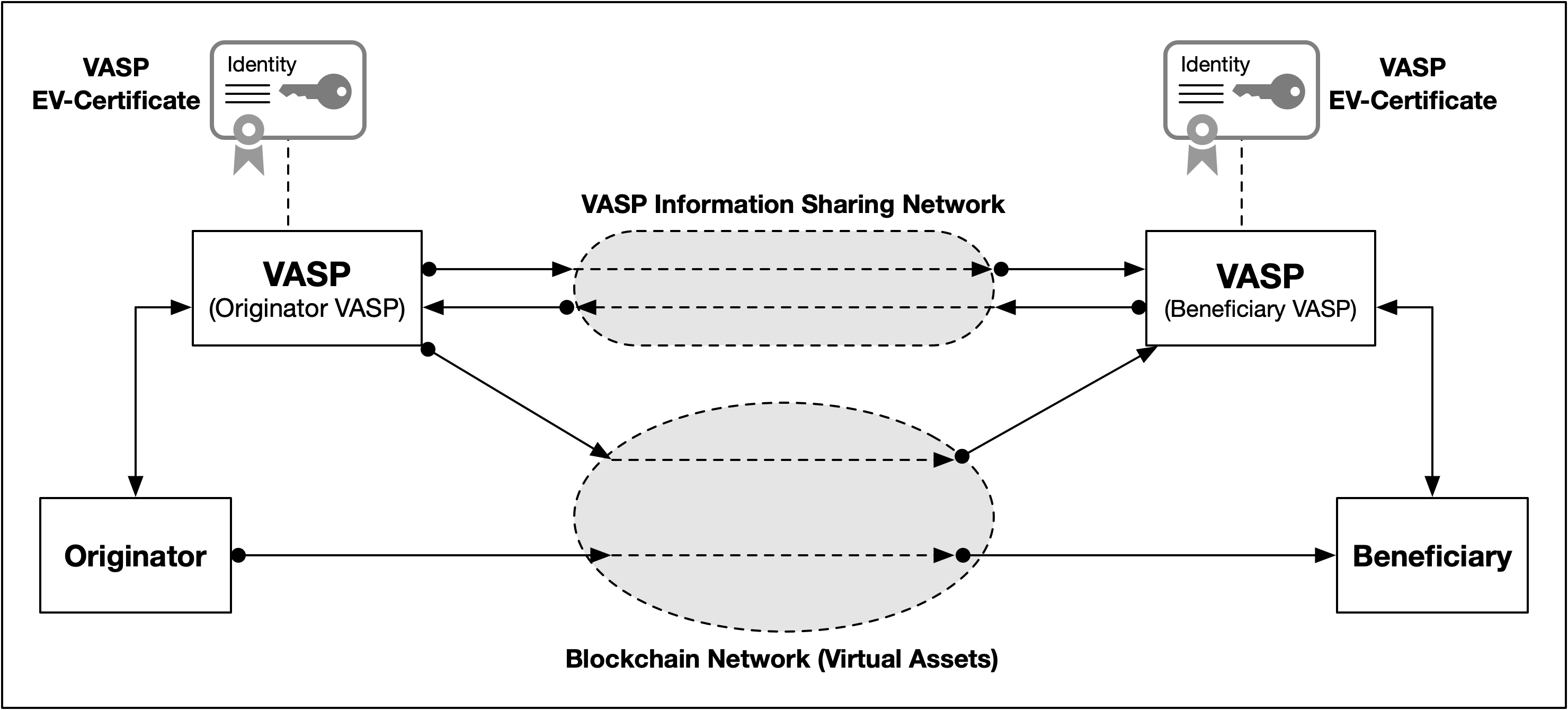}
\caption{Overview of VASP Information Sharing  Network (after~\cite{Hardjono2019b,TRISA2020-v07})}
\label{fig:VASPMessagingNetwork}
\end{figure}

There are several fundamental requirements for
an information sharing network for VASPs (Figure~\ref{fig:VASPMessagingNetwork}):
\begin{itemize}

\item	{\em Security, reliability and confidentiality of transport}:
The VASP information sharing network must provide 
security, reliability and confidentiality of communications between
an Originator-VASP and Beneficiary-VASP.
Several standards exists to fulfil this requirement
(e.g. IPsec VPNs~\cite{RFC2401-formatted}, 
{TLS} secure channels~\cite{RFC8446-formatted}, etc.).


\item	{\em Strong end-point identification and authentication}:
VASPs must use strong endpoint identification and authentication mechanisms
to ensure source and destination authenticity
and to prevent (reduce) man-in-the-middle types of attacks.
Mechanisms such as {X.509} certificates~\cite{rfc5280,ISO9594-pubkey} have been
used for over two decades across
various industries, government and defense
as a practical means to achieve this goal~\cite{Hardjono2019b}.

\item	{\em Correlation of customer information with on-chain transactions}:
There must be a mechanism to permit a VASP to accurately correlate (match)
between customer information (exchanged within the VASP information sharing network) 
and the blockchain transactions belonging to the respective customers.
This must be true also in the case of {\em batch transactions} performed by a VASP
(e.g. in the commingled accounts business model).

\item	{\em Consent from originator and beneficiary for customer information exchange}:
Unambiguous consent~\cite{GDPR} must be obtained by VASPs
from their customers with regards to the 
transmittal of customer personal information to another VASP.
Explicit consent must also be obtained from the beneficiary
for receiving asset transfers from an originator.
That is, a Beneficiary-VASP must obtain consent from its customer
to receive asset transfers into the customer's account.

\end{itemize}
Efforts are currently underway to begin 
addressing the need for a VASP information sharing network
to support VASPs in complying to the various aspects of 
the Travel Rule (see~\cite{TRISA2020-v07,OpenVASP-2019}).
A standard customer information model has recently been developed~\cite{InterVASP2020FINAL} 
that would allow VASPs to interoperate with each other with semantic consistency.

\section{A Trusted Identity Infrastructure for VASPs}
\label{sec:TrustedIdentityVASPs}

Another part of the information sharing infrastructure
is a VASP {\em trusted identity infrastructure}
that permits VASPs to prove their identity, public-key(s),
and legal business information.

A trusted identity infrastructure must address the various
challenges around VASP identities and provide the following types of mechanism:
\begin{itemize}

\item	{\em Discovery of VASP identity and verification of business status}:
Mechanism are needed to permit any entity on the Internet 
to ascertain whether a virtual asset service provider is a regulated VASP
within a given jurisdiction. 
An Originator-VASP must be able to easily locate the identity information for Beneficiary-VASP 
and to rapidly determine 
the business and legal status of that Beneficiary-VASP (vice versa).

\item	{\em Discovery and verification of VASP public-keys}:
Mechanism are needed to permit any entity on the Internet 
to ascertain whether a given public-key legally belongs to (operated by) a given VASP.

\item	{\em Discovery and verification of VASP service endpoints}: 
Mechanisms are needed to permit a VASP to ascertain whether 
it is connecting to the  legitimate service endpoints (e.g. URI)
of another VASP
(and not a rogue endpoint belonging to an attacker).

\item	{\em Discovery of VASPs using customer identifiers}:
Mechanisms are needed to permit a VASP to search and discover
a binding (association) between a customer 
user-friendly identifier (e.g. email address)
and the VASP (one or more) that may hold an account
for that customer.

\end{itemize}

\subsection{Extended Validation Certificates for VASPs Business Identity}
\label{subsec:EVCertificates}

The problem of discovering and verifying service provider public-keys and service endpoints
was faced by numerous online merchants nearly two decades ago.
For the end-user (i.e. home consumer) it was increasingly difficult to distinguish
between a legitimate service provider (e.g. online merchant)
from rogue web-servers that mimic the look-and-feel of legitimate merchants.
In response to a growing trend of man-in-the-middle attacks,
a number of browser vendors established an alliance in the late 2000s 
-- called the {\em CA/Browser Forum} (CAB Forum) -- that brought together
browser vendors and X.509 certification authorities (CA).
The CAB forum, as an industry standards defining organization,
published a number of industry technical specifications 
referred to as {\em Extended Validation} (EV) identity certificates~\cite{CAB-Forum2020}.
The overall goal was to enhance the basic X.509 certificate~\cite{rfc5280,ISO9594-pubkey}
with additional business related information regarding the subject (i.e. the online merchant).
The CA that issues EV-certificates must perform the various
information background checks regarding the subject,
to ensure that the subject was a legitimate business.
Correspondingly, the browser vendors supported EV-certificates
by pre-installing a copy of the root CA certificate
of all compliant CAs into their browser software.

We believe a similar approach is suitable for
fulfilling a number of the VASP requirements discussed above.
Some of the subject (VASP) business information 
to be included in the VASP identity EV-certificate
could be as follows~\cite{TRISA2020-v07}:
\begin{itemize}

\item	{\em Organization name}: The Organization field must contain the 
full legal name of the VASP legal entity controlling the VASP service endpoint,
as listed in the official records in the VASP's jurisdiction.

\item	{\em VASP Alternative Name Extension}: 
The Domain Name(s) owned or controlled by the VASP 
and to be associated with the VASP's server 
as the endpoint associated with the certificate.

\item	{\em VASP Incorporation Number or LEI} (if available): 
This field must contain the unique Incorporation Number assigned 
by the Incorporating Agency in the jurisdiction of incorporation. 
If an LEI number~\cite{GLEIF2018} is available, then the LEI number should be used instead.

\item	{\em VASP Address of Place of Business}: 
This is the address of the physical location of the VASP's Place of Business.

\item	{\em VASP Jurisdiction of Incorporation or Registration}: 
This field contain information regarding 
the Incorporating Agency or Registration Agency.

\item	{\em VASP Number}: This is the globally unique VASP number, if used
(see OpenVASP~\cite{OpenVASP-2019}).

\item	{\em VASP Regulated Business Activity}: Currently,
no formal definition of business activity specific for VASPs have been defined.
Note that in reality
VASPs may operate different functions in the virtual assets ecosystem
(e.g. crypto-exchanges, virtual asset based fund managers, stablecoin issuers, etc.).

\item	{\em EV Certificate Policy Object Identifier}: 
This is the identifier for the policies that determine
the certificate processing rules.
Such policies could be created by the organization using the certificate,
such a consortium of VASPs (see section~\ref{subsec:Consortium-Model} below).

\end{itemize}

\subsection{VASP Transactions-signing and Claims-signing Certificates}

For assets in commingled accounts managed by a VASP,
the asset transfer on the blockchain is performed
by the VASP using its own private-public key pair on behalf of the customer.
The customer holds no keys in the commingled cases.
We refer to these private-public keys as the VASP
{\em transactions signing-keys}, and we refer to
corresponding certificate as the {\em transactions signing-key certificates}. 
The purpose of signing-key certificates is to certify the ownership of 
the private-public keys as belonging to the VASP.
A given VASP may own multiple transactions signing-keys 
and therefore multiple signing-key certificates.

Because a VASP must stand behind the customer information
it provides to other VASPs,
any {\em claims}~\cite{Sporny2019} or 
{\em assertions}~\cite{SAMLcore} that a VASP produces
about its customers must be digitally signed by the VASP.
We refer to these private-public keys as 
{\em claims signing-keys}, and we refer to
corresponding certificate as the {\em claims signing-key certificates}.

It is crucial for a VASP that these three (3) key-pairs be distinct. 
This is because the key-usage purpose of the keys are different,
and each key may have differing lifetime durations.
Depending on the profile of a transactions signing-key certificate
and the claims signing-key certificate,
they may include the serial number (or hash) of the identity EV-certificate of the VASP.
This provides a mechanism for a recipient
to validate that the owner of these two certificates
is the same legal entity as the owner of the VASP identity EV-certificate.

\subsection{Consortium-based VASP Certificate Hierarchy}
\label{subsec:Consortium-Model}

In order for VASPs to have a high degree of interoperability
-- at both the technical and legal levels --
a {\em consortium} arrangement provides a number of advantages
for the information sharing network.
Members in a consortium are free to collectively define 
the common {\em operating rules} that members must abide by.
The operating rules become input matter into the definition of the {\em legal trust framework}
which expresses the contractual obligations of the members.

A well-crafted set of operating rules 
for a VASP information sharing network provides its
members with the several benefits.
First,
it provides a means for the members to {\em improve risk management} 
because the operating rules will allow members to quantify and manage risks
inherent in participating within the network.
Secondly, the operating rules provides  
its members with {\em legal certainty and predictability}
by addressing the legal rights, responsibilities, and liabilities of participating within the network.
Thirdly, the operating rules provides
{\em transparency} to the members of the network 
by having all members agree to the same terms of membership (i.e. contract).
Since the operating rules is a legal contract,
it is legally enforceable upon all members.
Finally,
a set of operating rules which define common 
technical specifications (e.g. APIs, cryptographic functions, certificates, etc.), 
for all the members
provides the highest chance of technical interoperability of services.
In turn, this reduces overall system-development costs by allowing entities to re-use
implementations of those standardized technical specifications.
Several examples of consortium-based operating rules exist today
(e.g. NACHA~\cite{NACHA2019}, OIX~\cite{OIX2017}).

\begin{figure}[t]
\centering
\includegraphics[width=1.0\textwidth, trim={0.0cm 0.0cm 0.0cm 0.0cm}, clip]{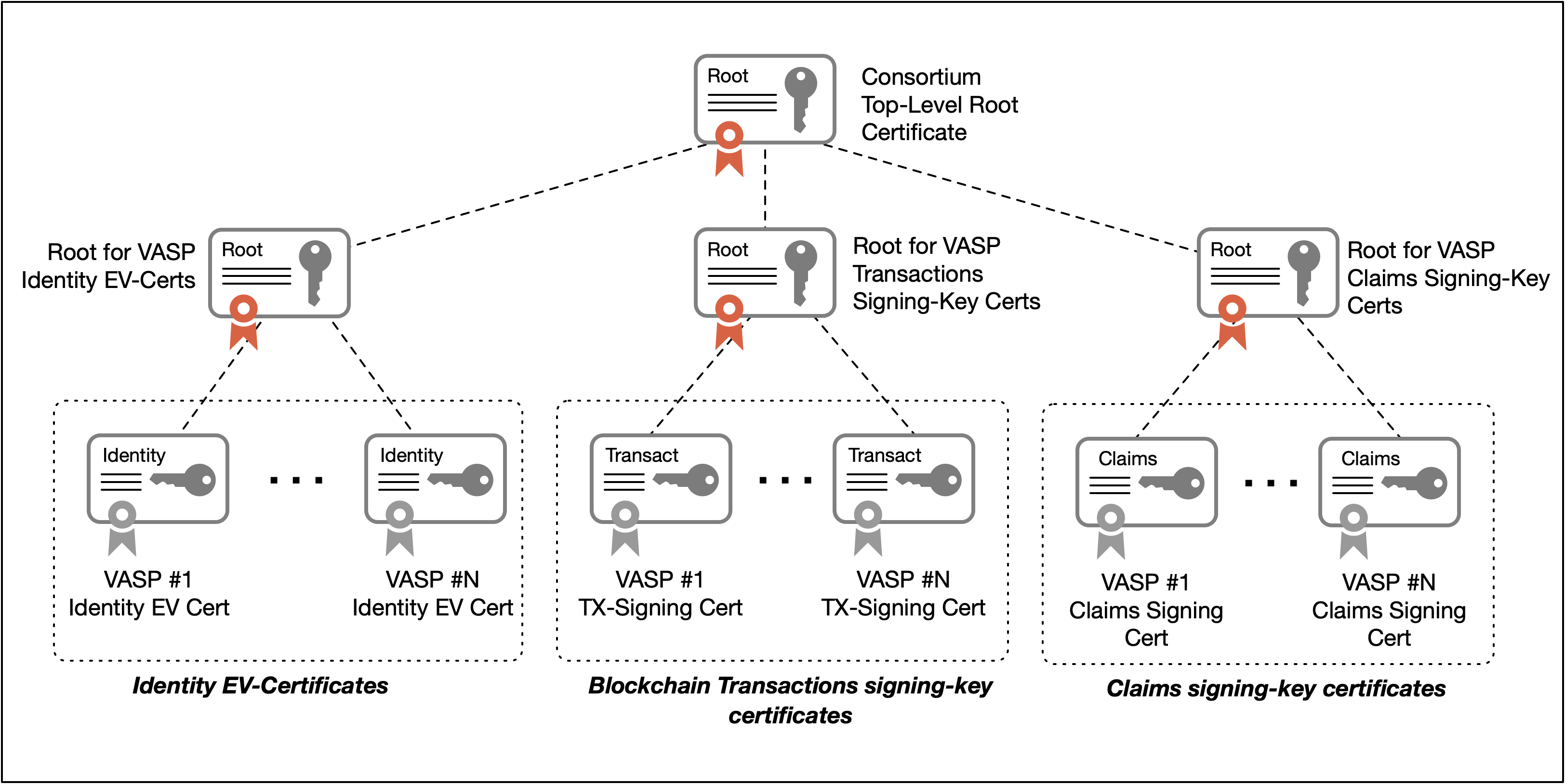}
\caption{Overview of a certificate hierarchy for VASP consortium}
\label{fig:keyhierarchy}
\end{figure}

Using the identity EV-certificates mentioned above
as an example,
the common operating rules would define the technical specification (profile)
of the EV-certificate 
(e.g. cryptographic algorithms, key lengths, duration of validity,
issuance protocols, revocation protocols, etc.),
as well as the legal information that must be included in the EV fields
of the certificate
(e.g. legal incorporation number, LEI number, place of business, etc.).

Business interoperability can only be achieved if all members of the information sharing network
observe and implement these common operating rules,
and if there is legal and monetary liability for not doing so.
This approach is not new and is used for group peering agreements 
among IP routing service providers (i.e. access ISPs and backbone ISPs).
Technological interoperability of identity EV-certificates
dictates that members of the information sharing network
participate under a common {\em certificate hierarchy},
that is rooted at the consortium organization.
This is shown in Figure~\ref{fig:keyhierarchy}, where the consortium 
becomes the Root-CA for the certificates issued 
to all VASPs in the consortium organization.

Certificate hierarchies have been successfully deployed in numerous
organizations,
ranging from government organizations~\cite{NIST-800-32},
to financial networks~\cite{SwiftNet2004},
to mobile devices and networks~\cite{AppleCPS2019},
to consortiums of cable-device manufacturers~\cite{CableLabsPKI2019}.
An example of a consortium that brings together 
device manufacturers (i.e. Cable Modem and Set Top Box vendors)
and service operators (i.e. regional cable access providers)
is Cable Laboratories (CableLabs).
The combined type of membership in CableLabs
permits cable operators to detect and isolate counterfeit devices
and provide end-to-end protection for valuable content (e.g. movies).
This combined approach may be relevant for VASPs dealing with customer wallet devices.

\section{Customer Identity and Key Management Infrastructure}
\label{sec:IdentityManagementInfra}

There is today a need for crypto-asset management systems
to be integrated seamlessly with existing identity management infrastructure functions,
including identity authentication services, authorization services, and
consent management services.

\begin{figure}[t]
\centering
\includegraphics[width=1.0\textwidth, trim={0.0cm 0.0cm 0.0cm 0.0cm}, clip]{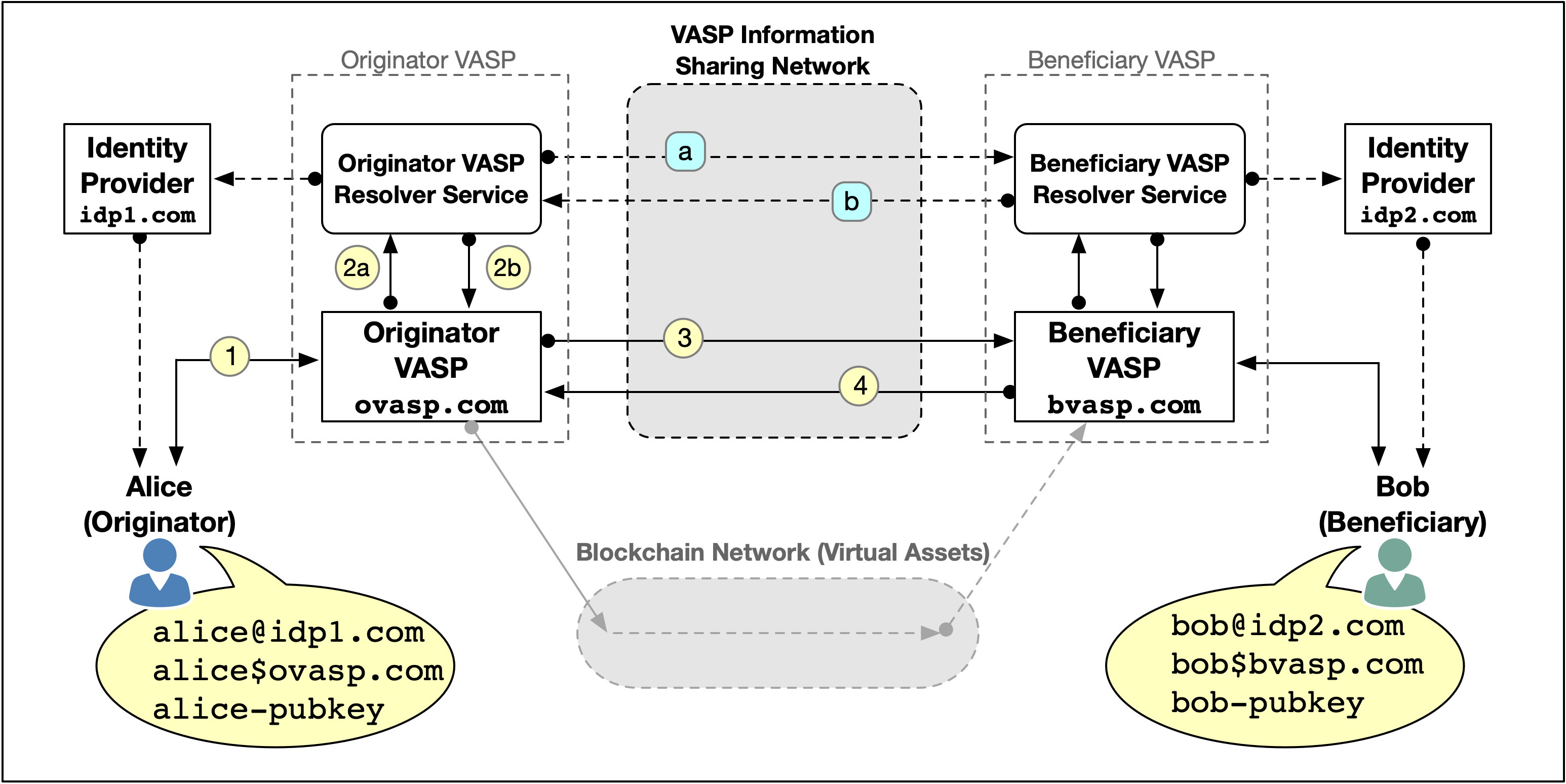}
\caption{Alice and Bob with multiple identifiers}
\label{fig:resolution-services}
\end{figure}

\subsection{Customer Identities and Digital Identifiers}
\label{subsec:Customer-Identifiers}

Currently most users employ their email address
as a form of user {\em identifier} in the context of obtaining services on the Internet.
Many of these identifiers do not represent 
the user's full person (core) identity~\cite{HardjonoPentland2019Book-CoreID},
and have short-term or ephemeral value (i.e. identifier can be replaced with a new one).
Typically the entities who issues the identifiers are 
email providers and social media platform providers.
The industry jargon used to describe them (rather inaccurately)
is {\em Identity Provider} (IdP).
Besides providing email-routable identifiers to users,
the identity provider's role in the identity ecosystem is to provide
mediated authentication services and credential-management 
on behalf of the user~\cite{Hardjono2019-IEEECommsMagazine}.
Mediated authentication -- such as single sign-on (SSO) -- provides
convenience for the user by obviating the need for them
to authenticate multiple times to each online service provider
(e.g. online merchant) they visit.
The online merchants redirects the user temporarily to
the IdP for user-authentication,
and upon success the user is returned back to the merchant.

The predominance of email-identifiers for users is a matter of consideration
for VASPs because some users may wish to use their email address
as the main identifier for account-creation at the VASP.
They may also seek to use the email identifier of a beneficiary
in the context of asset transfers.
Furthermore,
a user may have multiple accounts, each at different VASPs and each employing
different email identifiers obtained from different IdPs.

An interesting approach is used in the PayID scheme~\cite{PayID-2020-v1} 
where the user is identified using a string similar to the
{\em addr-spec} identifier (RFC5322), 
but with the ``@'' symbol replaced by the dollar sign (``\$'')
while retaining the local-part and the domain-part.
For example,
if Alice has an account at the PayID provider (e.g. ACMEpay.com)
then her PayID identifier would be
{\tt alice\$acmepay.com}.

\subsection{Identifier Resolvers}
\label{subsec:PKResoltion}

The matter of user identifiers is important 
not only from the customer usability perspective,
but also from the need for interoperability of services across VASPs
within the information sharing network.
When an Originator-VASP employs a user-identifier scheme to identify
an originator and beneficiary,
then the Beneficiary-VASP must have the 
same syntactic and semantic understanding of the identifier scheme.
That is, both VASPs must be identifying
the same pair of originator and beneficiary customers.
Thus, another use of the VASP information sharing infrastructure
discussed previously is
for VASPs to exchange the list of identifiers of their respective customers.
This can be achieved by each VASP in the network
employing an {\em Identifier Resolver Service} (server)
that is accessible to other VASPs in the network
(see Figure~\ref{fig:resolution-services}).

Some of the general requirements for a VASP resolver service
are as follows (non-exhaustive):
\begin{itemize}

\item	{\em Support for multiple user-identifiers}:
The resolver service must permit multiple types of user-identifiers
to be associated with the customer of the VASP.

\item	{\em Fast lookup for VASP determination}:
The resolver service must support fast look-ups or searches
based on an identifier string by other VASPs in the network.
Such look-ups may be part of an asset-transfer request
from a VASP's customer and any delays in identifying the Beneficiary-VASP
may add to the overall transfer settlement time
as perceived by the customer.

\item	{\em Protected service APIs}:
The service endpoint APIs (e.g. RESTful, PubSub, etc.)
of the resolver service must be protected.
A caller VASP must be authenticated and authorized to use the APIs.

\item	{\em Validation of user-identifier to IdP}:
Optionally,
for every user-identifier string submitted (added to) by a customer
to their account at the VASP,
the resolver service of the VASP should validate 
the string to its original issuer (if it was not the VASP).
Thus, if customer Alice wishes to employ
her email address {\tt alice@idp1.com} then the resolver service
should seek to validate that Alice is known to the provider IdP1.

\end{itemize}

Figure~\ref{fig:resolution-services} illustrates
that both Alice and Bob
can be recognized using three different means:
(i) their email address issued by an IdP (e.g. {\tt alice@idp1.com});
(ii) their PayID address managed by a VASP (e.g. {\tt alice\$ovasp.com});
or
(iii) their bare public-keys (e.g. {\tt alice-pubkey}).

Using Figure~\ref{fig:resolution-services},
consider the example of Alice who
wishes to transfer virtual assets to Bob,
but who only knows Bob's email address (e.g. {\tt bob@idp2.com}).
Alice does not know Bob's public-key or  Bob's VASP.
This means that Alice's Originator-VASP must
query its resolver service
-- as shown in Step~2(a) and Step~2(b) of Figure~\ref{fig:resolution-services} --
to discover which other VASPs in the network may know of the string {\tt bob@idp2.com}
(i.e. uses the string in an account).
Assuming the resolver service returns a positive response
(i.e. VASP-identifier or VASP-number found),
the Originator-VASP can begin inquiring to that VASP about Bob per the Travel Rule,
as summarized in Step~3 and Step~4 of Figure~\ref{fig:resolution-services}.

Note that the resolver service of the Originator-VASP
may return more than one possible Beneficiary VASP-identifier or VASP-number.
This could mean that Bob has an active account at each of these VASPs
(each possibly using a different private-public key pair).
In such cases,
the Originator-VASP may need to request further information (regarding Bob)
from Alice.

It is worth noting that
that identifier resolvers are not new,
and several resolver protocols have been standardized 
and have been in wide deployment for over two decades now
(e.g. Domain Name Service (RFC1035), Handle System (RFC3650), etc.).
As such, the nascent VASP industry should consider using and extending these
well-deployed systems, instead of designing something from scratch.

\subsection{Customer Privacy and Resolving to Public-Keys}
\label{subsec:CustomerPrivacy}

In general we believe that a VASP resolver service
should not return customer public-keys in the first instance
in order to preserve customer privacy.

The purpose of the VASP resolver service
is to aid other VASPs in determining whether an entity (person or organization)
is a customer of one (or more) of the VASPs in the network.
That is,
the resolver service is aimed at solving the VASP {\em discoverability} problem:
namely to look-up VASP identifiers (VASP-numbers)
in order to engage that VASP.

Secondly, the information (metadata) about
the association between a user-identifier and a VASP
is less revealing than the association between a user-identifier and a public-key.
As we mentioned before,
user-identifiers (e.g. email addresses) associated with a customer account at VASP
can be changed by the customer at anytime without impacting
the virtual assets bound to the customer's public-key.
In contrast,
a change to the customer's public-key is visible on the blockchain system.

Finally,
different VASPs may employ different business models
(e.g. key-custodian, commingled funds (accounts-only), regulated customer wallets).
As such,
in some cases (commingled funds) there is in fact no unique public-key
associated with a given customer.

\subsection{Federation of Resolvers: VASP Information Sharing Network}
\label{subsec:FederationResolvers}

In order for VASP resolver services to scale-up,
the VASPs must {\em federate} their resolver services
under a common legal framework 
(i.e. the consortium model discussed above).
A federation agreement allows
VASPs to share customer identifier information
(as known to the VASP)
over the information sharing network 
(discussed in Section~\ref{sec:GlobalVASPMessagingNetwork}).
Indeed, this is one of the main purposes of the network.

For example,
using the information sharing network, 
VASPs can regularly (e.g. overnight)
exchange knowledge about each other's customer identifiers.
This is shown in Figure~\ref{fig:resolution-services}
in Step~(a) and Step~(b) that runs between the VASP resolver services.

Although the precise protocol is beyond the scope of the current work,
in the simplest form the exchange of customer identifier information
between VASP resolver services
can consists of pairs of VASP-identifier value and customer-identifier values
(i.e. list of customer-identifiers as known to the VASP):
\begin{center}
{\tt VASP-identifier}, {\tt Customer-id-1}, {\tt Customer-id-2}, $\dots$ {\tt Customer-id-N}
\end{center}
This approach is akin to IP route Link State Advertisements (LSA) used
within some link-state routing protocols
(e.g. OSPF (RFC2328)).
In this case, a VASP is ``advertising'' its knowledge of customers bearing the stated identifiers.

The exchange of customer identifier information
between two VASP resolver services must be conducted
through a SSL/TLS secure channel established 
using the VASP {X.509} EV-certificates
to ensure traffic confidentiality and source authenticity.


\subsection{Customer Managed Access to Claims}
\label{subsec:customerUMA}

In some cases,
static attributes regarding a customer (e.g. age, state of residence, driving license, etc)
can be obtained from authorized entities (e.g. government departments)
in the form of asserted claims in one format or another~\cite{SAMLcore,Sporny2019},
with a fixed validity period.
Within the identity industry,
entities which issue signed assertions or claims are referred to as the {\em Claims Provider} (CP).
%
A given customer of a VASP may already be in possession of 
a signed claim (e.g. driving license number)
from an authoritative CP (e.g. Department of Motorized Vehicles),
where the claim signature is still valid.
The customer may keep a copy of the signed claim 
in the customer's personal {\em Claims Store}
(e.g. mobile device, home server, cloud storage, etc.).
The customer could provide its VASP with a copy of the claims,
or the customer may provide its VASP with access to the customer's claims store.

\begin{figure}[t]
\centering
\includegraphics[width=1.0\textwidth, trim={0.0cm 0.0cm 0.0cm 0.0cm}, clip]{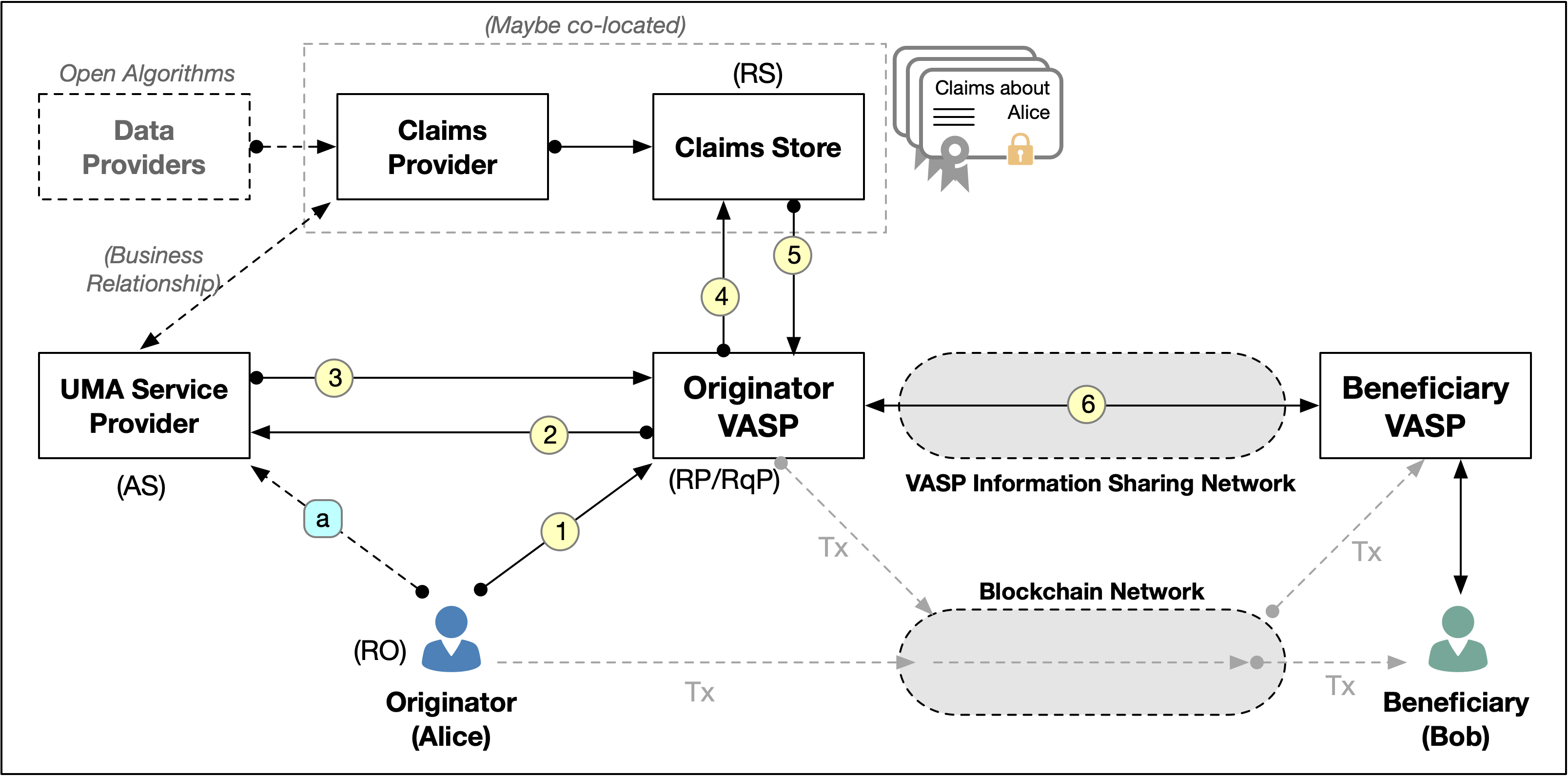}
\caption{Overview of originator authorization for VASP to retrieve claims}
\label{fig:claimsgathering}
\end{figure}

There are several requirements and challenges for VASP access to claims store managed by the customer:
\begin{itemize}

\item	{\em Customer-managed authorization to access claims-store}:
Access to the customer's claims-store must be customer driven,
where access policies (rules) are determined by the customer as the claims owner.

\item	{\em Notice and consent from customer to use specific claims}:
The customer as the claims owner must determine via access policies 
(i) which claims are accessible (readable) to the VASP,
(ii)  the usage-purpose limitations of the claims,
(iii) and the right for the customer to retract (withdraw) the consent~\cite{GDPR}.
The customer's claims store 
must provide notice to the VASP, and the VASP must agree to the terms of use.

\item	{\em Consent-receipt issuance to VASP}:
The customer's claims-store must issue a {\em consent-receipt}~\cite{CONSENT1.0}
to the VASP
which acts exculpatory evidence covering the VASP.

\end{itemize}
An extension to the {OAuth2.0} framework~\cite{rfc6749} called 
the {\em User Managed Access} (UMA) protocol~\cite{UMACORE1.0,UMACORE2.0}
can be the basis for customer-managed access to the claims store.
In the example of Figure~\ref{fig:claimsgathering},
the Originator-VASP is seeking to obtain signed claims
regarding the originator customer (Alice) located in her claims store.
Access polices have been set by Alice in Step~(a).
In Step~1 Alice provides her VASP with the location of this service provider.
When the VASP reaches the UMA Service Provider (Step~2) -- which acts as
the authorization server in the {OAuth2.0} and UMA context --
the VASP is provided with authorization-token
that identifies the specific claims the VASP is authorized to fetch (Step~3).
The VASP wields the authorization-token
to the claims-store (Step~4).
The VASP obtains access to the relevant claims and is provided
with a consent-receipt by the claims-store (Step~5).

A claim-store can be implemented in several ways.
For example,
it can be a Resource Server under the control of the CP,
it can reside on Alice's own mobile device,
it can be placed in a cloud-based 
Trusted Execution Environment (TEE)~\cite{EEA2019-Trusted-Compute},
or it may implemented in a decentralized 
file system such as IPFS/Filecoin~\cite{IPFS} based on 
a decentralized identifier (DID) scheme~\cite{W3C-DID-2018}.

\section{Attestations Infrastructures for Regulated Wallets}
\label{sec:AttestationsInfrastructures}

With the increase in the number of individuals and organizations 
holding private-public keys bound to virtual assets on a blockchain,
there is an increased risk of the loss and/or theft of private-keys.
VASPs who are custodians of a customer's private-public keys 
and VASPs who employ their own keys to transact on behalf of customers
face the problem of key management.
As such,
the use of  {\em electronic wallets} based on {\em trusted hardware}
-- such as the Trusted Platform Module (TPM) chip~\cite{TPM1.2specification} --
that offers key-protection
may increasingly become a necessity for VASPs for their own business survival.
Funds insurance providers~\cite{KharifLouis2018} may seek to obtain evidence
of the use of trusted hardware by VASPs and their customers. 
This brings to the foreground the challenge of establishing an {\em attestation infrastructure}
for VASPs that assists them in obtaining greater visibility 
into the state of wallets implemented using trusted hardware.

In the following we use the term {\em regulated wallet}  
to denote a wallet system (hardware and software) that is in possession
of a customer of a supervised (regulated) VASP~\cite{FINMA-Guidance-2019,AMLO-FINMA2015}.
We use the term {\em private wallet} to denote
a wallet system belonging to an {\em unverified entity}~\cite{FATF-12MonthReview-2020}.
In some cases a VASP may decline to perform an asset-transfer
to a public-key thought to be controlled by a wallet simply because
the wallet-holder information is unattainable by a VASP,
despite the VASP querying other VASPs in the information sharing network.

\subsection{Attestation Information Relevant to VASPs \& Asset Insurers}

We use the term {\em attestation}
to mean the capability in some trusted hardware
to provide proof 
that a device (platform) using the trusted hardware
can be trusted to correctly and truthfully report
the internal state of the device~\cite{TCG-Glossary-2017}.
The information reported is signed by the trusted hardware
using an internal private-key that is non-readable by external entities
and {\em non-migrateable} (i.e. cannot be extracted) from the trusted hardware.
Thus, the recipient (e.g. a verifier) of the evidence 
obtains assurance that the signed report came from 
a particular device with the specific 
trusted hardware~\cite{TCG-Integrity-Schema-2006,TCG-Attestations-Arch2020}.
These features provides interesting capabilities for VASPs
in addressing some of their key management challenges,
as well as AML/FT compliance needs.

For VASPs and asset insurance providers,
there are several types of {\em attestation evidence} information that can be obtained
from a wallet regarding keys used 
to sign asset-related transactions on the blockchain.
The type of attestation evidence is dependent
on the specific type trusted hardware but generally
consists of the following~\cite{HardjonoLipton2020-wallet-arxiv}:
\begin{itemize}

\item	{\em Key creation evidence}:
The trusted hardware used in a wallet must have the capability
to provide evidence regarding the origin of cryptographic keys held by the hardware.
More specifically,
it must be able to attest as to whether it generated a private-public key-pair internally
or whether the key-pair was imported from outside.

\item	{\em Key-movability evidence}:
The trusted hardware used in a wallet must have the capability
to provide evidence as to whether a private-key is 
migrateable or non-migrateable~\cite{HardjonoKazmierczak2008}.
This evidence permits the VASP to perform risk-evaluation regarding the possibility
that the wallet holder (i.e. its customer) has dishonestly exported a copy of private-key 
to another wallet (and then use the key-pair in an unregulated manner).

\item	{\em Wallet system stack evidence}:
The trusted hardware used in a wallet should have the capability
to provide evidence of the software stack present 
in the wallet~\cite{TCG-Integrity-Schema-2006,TCG-Attestations-Arch2020}.

\end{itemize}

A given VASP may demand that customers use only approved wallets
based on suitable trusted hardware.
The VASP may also demand customers to create and use 
new key-pairs in the trusted hardware for all transactions
(i.e. from the time the user becomes a legal customer of the VASP).
This strategy provides the VASP with a clear line of responsibility and accountability
under the Travel Rule with regards to customer-originated transactions.
The VASP has exculpatory evidence regarding the on-boarding of the new customer
and the start of use of the new key-pair.

Virtual asset insurance providers (e.g. crypto-funds insurers)
must have the ability to directly query wallets in order to obtain
signed attestation evidence from the trusted hardware in the wallets. 
This gives funds-insurance providers with visibility 
regarding the robustness of the key management lifecycle employed by the VASPs
for their regulated wallets,
and provides them with tangible information upon which to make their risk-assessment.

\subsection{On-Boarding and Off-Boarding Customers}
\label{subsec:OffBoarding}

There are a number of challenges related to the on-boarding of a customer
already possessing a wallet.
In the case that the customer wallet is regulated and previously known to another regulated VASP,
then there are some practical considerations
that the {\em acquiring} VASP needs to address.
These include:
(i) validating whether prior to on-boarding the wallet was regulated or private;
(ii) validating that the keys present within the wallet corresponds to the customer's
historical transactions (confirmed on the blockchain);
(iii) verifying whether a backup/migration of the wallet has occurred in the past;
(iv) determining whether the customer's assets should be moved to new keys, and if so,
how the ``old'' keys will be treated;
and so on.

The case of a customer leaving a VASP (i.e. off-boarding) also introduces a number
of questions that may be relevant under the Travel Rule.
The {\em releasing} VASP may need to address the following:
(i) preparing evidence that the wallet was in a regulated state whilst
the owner of the wallet was a customer of the VASP;
(ii) whether the customer's assets should be moved to a temporary set of keys,
denoting the end of the VASP's responsibilities for the customer under the Travel Rule;
(iii) obtaining evidence from the wallet that the ``old'' keys (non-migrateable keys)
have been erased from the wallet device,
thereby rendering the keys unusable in the future by the customer;
and so on.

\section{Conclusions}
\label{sec:Conclusions}

The VASP information sharing network is a core component
of the trust infrastructures needed if blockchain systems and virtual assets
are to be the foundation of the future global digital economy.
VASPs need to view this information sharing network
as a foundational building block for other infrastructures to be developed.

VASPs also require a trusted identity infrastructure
that allow VASPs to authenticate each other 
and to rapidly ascertain the business legal status of other VASPs.
The use of extended-validation digital certificates offers
a promising solution to this problem,
based on well understood and widely deployed public key certificates management technologies.

Finally,
other trust infrastructures will be needed in order to address
use-cases related to customer wallets and device attestations from wallets.
In particular,
VASPs may need evidence that the customer's 
private-key truly resides within the wallet device.
This provides a means for VASPs to prove that they are not the
legal operator of the customer's private-public keys.

\section*{Acknowledgements}

We thank the following for various inputs, discussions and comments:
Sandy Pentland and Alexander Lipton (MIT);
Anne Wallwork (US Treasury);
Dave Jevans, John Jefferies, Frank Steegmans (CipherTrace);
David Riegelnig (Bitcoin Suisse);
Aanchal Malhotra (Ripple);
Justin Newton (NetKi);
Eve Maler (ForgeRock);
Justin Richer (Bespoke Engineering);
Nat Sakimura (OIF).



\end{document}